\documentclass[aps,pre,twocolumn,nofootnoteinbib,superscriptaddress]{revtex4-2}
\usepackage{graphicx,amssymb,amsmath}
\usepackage{color}
\usepackage{soul}
\usepackage[dvipsnames]{xcolor}

\newcommand{\ff}[2]{\includegraphics[width=#1\columnwidth]{fig.current/#2}}

\newcommand{\ave}[1]{\langle {#1} \rangle}
\newcommand{\pas}[1]{\left[ {#1} \right]}

\newcommand{\Coop}{{\rm C}}
\newcommand{\Defect}{{\rm D}}

\newcommand{\boundary}{{x}}
\newcommand{\fixation}{{\rm f}}
\newcommand{\crit}{{\rm crit}}
\newcommand{\threshold}{{\rm th}}
\newcommand{\theoretical}{{\rm (theo)}}
\newcommand{\occupation}{{\rm (occu)}}
\newcommand{\nowealth}{{\rm (nw)}}

\begin{document}
 
\title{Discontinuous Wealth-Gradient Transition Driving Cooperation}
\author{Hyun Gyu Lee}
\affiliation{Department of Physics and Astronomy, Sejong University, Seoul, 05006, Korea}
\affiliation{School of Computational Sciences, Korea Institute for Advanced Study, Seoul 02455, Korea}
\author{Hyeong-Chai Jeong}
\email{hcj@sejong.ac.kr}
\affiliation{Department of Physics and Astronomy, Sejong University, Seoul, 05006, Korea}
\affiliation{School of Computational Sciences, Korea Institute for Advanced Study, Seoul 02455, Korea}
\author{Deok-Sun Lee}
\email{deoksunlee@kias.re.kr}
\affiliation{School of Computational Sciences, Korea Institute for Advanced Study, Seoul 02455, Korea}
\affiliation{Center for AI and Natural Sciences, Korea Institute for Advanced Study, Seoul 02455, Korea}

\date{\today}

\begin{abstract}
  The universal prevalence of cooperation is puzzling,
  as defection typically yields higher payoffs than cooperation,
  motivating searches for hidden pathways to cooperation.
  Here we study a game-theoretic model on a lattice structured population in which interaction payoffs are scaled by the minimum of participants' accumulated wealth, reflecting real-world heterogeneity and incorporating the influence of past strategic choices. This wealth scaling allows frequent cooperators to surpass defectors in payoffs through their greater wealth even at high cooperation costs where defection would otherwise dominate. At the elevated critical cost-benefit ratio, the wealth gradient at the cooperator-defector boundary in one dimension exhibits a discontinuous transition. We show that slowing and effective stalling of the boundary trigger an explosive buildup of the wealth gradient, driving the dominance of cooperation below the critical ratio. Remarkably, this promotion of cooperation is stronger at higher temperatures, revealing a constructive role of fluctuations.
\end{abstract}

\maketitle

\section{Introduction}
\label{sec:intro}

Cooperation, in which individuals incur a personal cost to benefit others,
is ubiquitous across biology and society, yet its persistence remains a central
puzzle~\cite{hamilton1964genetical,trivers1971evolution,fehr2003nature,nowak2006five}. Game theory formalizes this puzzle by showing that in social dilemmas individually rational choices lead to mutual defection, as in arms races and the tragedy of the
commons~\cite{schelling1960strategy,hardin1968tragedy,smith1982evolution}.
This apparent contradiction has motivated extensive studies on the emergence and maintenance of
cooperation~\cite{axelrod1981evolution,henrich2004cultural,szabo2007evolutionary,nowak1998indirect}. Foundational mechanisms proposed to explain cooperation include kin selection, direct reciprocity, indirect reciprocity, and network reciprocity~\cite{milinski2002reputation,nowak1992spatial,hauert2002volunteering,wilson1994group,traulsen2006multilevel}. Research on its evolution continues to expand across diverse settings~\cite{Gardenes2007, Wang2017,korolev2015evolution,su2022evolution,colnaghi2023adaptations,kleshnina2023effect,civilini2024explosive,sadekar2024evolutionary,garcia2025picking,meng2025promoting,tarnita2025reconciling}.

Cooperation in these mechanisms relies on the assortment of cooperators~\cite{FletcherDoebeli2009Assortment, NowakTarnitaAntal2010Structured, Taylor2007Transforming}, yet finite-temperature fluctuations can hinder this assortment~\cite{szabo2007evolutionary,li2021limited}. Given the ubiquity of both fluctuations and cooperation, how can cooperation arise through fluctuations? 
In this work, we elucidate a fluctuation-driven mechanism for cooperation  in a wealth-dependent game theoretic model~\cite{Chadefaux2010}, where payoffs are scaled by players' wealth, reflecting bounded, resource-proportional interactions. 
Individuals probabilistically adopt a neighbor’s current strategy, favoring those that achieved higher payoffs in the current round. Because these payoffs depend on players’ wealth, which accumulates through past interactions, competition between C and D clusters becomes history dependent.
In structured populations, clustered cooperators accumulate more wealth than defectors, generating a wealth gradient through which wealth-scaled payoffs can favor C over D clusters. 

Wealth transfer in real economic transactions has been modeled in various ways~\cite{patriarca_basic_2010}. A fundamental constraint is that traders do not risk losing more wealth than they own~\cite{hayes_computing_2002}.
This basic budget constraint motivates the yard-sale rule, where the amount of wealth transferred in an interaction is bounded by the minimum wealth of two participants~\cite{Chakraborti2002,hayes_computing_2002,boghosian_kinetics_2014,PhysRevE.108.064303}.
However, most game-theoretic models assume a fixed payoff scale that is independent of players’ wealth, whereas in real interactions the scale of a game typically depends on the resources of the participants. 
Chadefaux and Helbing~\cite{Chadefaux2010} addressed this issue by studying a game in which payoffs are proportional to the minimum wealth of the interacting players and showed that such wealth-scaled interactions promote cooperation at zero temperature.
Here, we extend this model to finite temperatures and derive analytic results in one dimension to illuminate cluster dynamics and the role of fluctuations.

In one dimension, we find that the wealth gradient undergoes a discontinuous transition at a critical cost-benefit ratio below which cooperation dominates. This critical point corresponds to a saddle-node bifurcation of a self-consistent equation for the wealth gradient, which is derived analytically in one dimension.  Above this value, a stable solution exists, indicating a persistent expansion of defector domains driven by their payoff advantage, leading to fixation to defection. In contrast, below the critical ratio the boundary between cooperator and defector domains moves non-monotonically. It initially moves to contract the cooperator domain, but the growing wealth gradient slows and effectively stalls the boundary,
giving rise to an explosive buildup of the wealth gradient at the boundary. 
The domain boundary then reverses direction, expanding the cooperator domain, and the system
fixates to cooperation. At higher temperatures, stronger fluctuations in  boundary motion amplify the wealth gradient and promote cooperation. In two dimensions, wealth-scaled interactions likewise enhance cooperation, with fluctuations strengthening this effect. Overall, our analysis of this simple model reveals the feedback between accumulated wealth shaped by past decisions and present decision-making dynamics, a mechanism that may underlie collective behaviors and decision-making in diverse contexts. 

\section{Model}
\label{sec:model}

We consider $N$ individuals, indexed by $i=1,2,\ldots, N$,
each adopting a strategy $s_i\in \{\rm C, D \}$ and possessing a wealth $W_i\geq 0$,
which is initially $W_i(0)=1$.
Between time steps $t$ to $t+1$, all individuals update their strategies in a random sequential
order without replacement.
As in the `death-birth' updating~\cite{ohtsuki_2006},  
the strategy of individual $k$ is replaced by that of one of its neighbors $i \in \mathcal{N}_k$, where $\mathcal{N}_k$ is the set of neighbors of $k$. 
These neighbors compete to replicate to $k$ according to the payoff $\sum_{j \in \mathcal{N}_i}{\Pi_{ij}^k}(t)$ from interactions with their own neighbors $j\in \mathcal{N}_i$.
Therefore, in individual $k$'s turn, each of its neighbors $i \in \mathcal{N}_k$ interacts with each of
its own neighbors $j\in\mathcal{N}_i$ to gain a payoff $\Pi_{ij}^k(t)$ defined as
\begin{align}
\label{eqn:payoff_pi}
\Pi_{ij}^k(t) 
& = \varepsilon \min\left[{W_i(t), W_j(t)}\right] (\delta_{s_j,\rm C}-c\, \delta_{s_i,\rm C}) \\
& = \varepsilon \min\left[{W_i(t), W_j(t)}\right]\begin{pmatrix} \notag
1-c & -c \\ 1 & 0
\end{pmatrix},
\end{align}
where $\varepsilon$ is a small constant and $c$ is the cost-benefit ratio in the range $0\leq c<1$,
given that the benefit is set to $1$. 
Note that $s_i$ and $s_j$ may either remain to be or may have already been updated depending
on the order of their update turns.
Individual $k$ adopts the strategy of its neighbor $i$ ($s_k \rightarrow s_k'=s_i$) with a probability that increases with the neighbor's total payoff, 
\begin{equation}
    P_{i}^k(t)=\frac{e^{\beta \frac{f_i^k(t)}{W_k(t)}}}{\sum_{\ell\in \mathcal{N}_k} e^{\beta \frac{f_\ell^k(t)}{W_k(t)}}} 
    \ {\rm with} \
     f_{\ell}^k(t)\equiv \sum_{j\in \mathcal{N}_\ell}\Pi_{\ell j}^k(t),
    \label{eqn:copy_probability}
\end{equation}
where $f_\ell^k(t)$ denotes the total payoff earned by individual $\ell$ from interactions with its
neighbors occurring during $k$'s update turn.
The inverse temperature $\beta$ sets the selection strength for strategies yielding higher payoffs, extending the zero-temperature limit ($\beta\to\infty$) studied in Ref.~\cite{Chadefaux2010}.
After all $N$ strategy updates, each individual's wealth is updated  by adding the payoffs
earned between $t$ and $t+1$ as
$
  W_i(t+1)
  = W_i(t) + \sum_{j\in \mathcal{N}_i}\left(\sum_{k\in \mathcal{N}_i} +\sum_{k\in \mathcal{N}_j}\right)
  \Pi_{ij}^k(t).
$  
We primarily focus on the one-dimensional case while also presenting two-dimensional results
in the later part.

When payoffs are independent of wealth ($W_i(t)=1$ for all $i,t$), consider a defector cluster  adjacent to a cooperator
cluster  ($s_i={\rm C}$ for $i\leq {\boundary}$ and $s_i={\rm D}$ for $i\geq {\boundary}+1$).
The cooperator at ${\boundary}$ is more likely to retain cooperation than switch to defection
if the payoff of its cooperating neighbor at ${\boundary}-1$,
$f_{{\boundary}-1}^{{\boundary}} = 2\varepsilon (1-c)$,
exceeds that of the defecting neighbor at ${\boundary}+1$, $f_{{\boundary}+1}^{{\boundary}}=\varepsilon$.
This condition holds if the cost-benefit ratio $c$ is below $c_\crit^\nowealth= 1/2$.
The same critical ratio arises from the condition for the defector at ${\boundary}+1$ is more
likely  to adopt cooperation.
Our main question is how the critical ratio shifts,
up (promoting cooperation) or down (suppressing cooperation),
when wealth becomes inhomogeneous and time-dependent, thereby influencing payoffs through Eq.~\eqref{eqn:payoff_pi}.  

\begin{figure}[t]
\ff{1.}{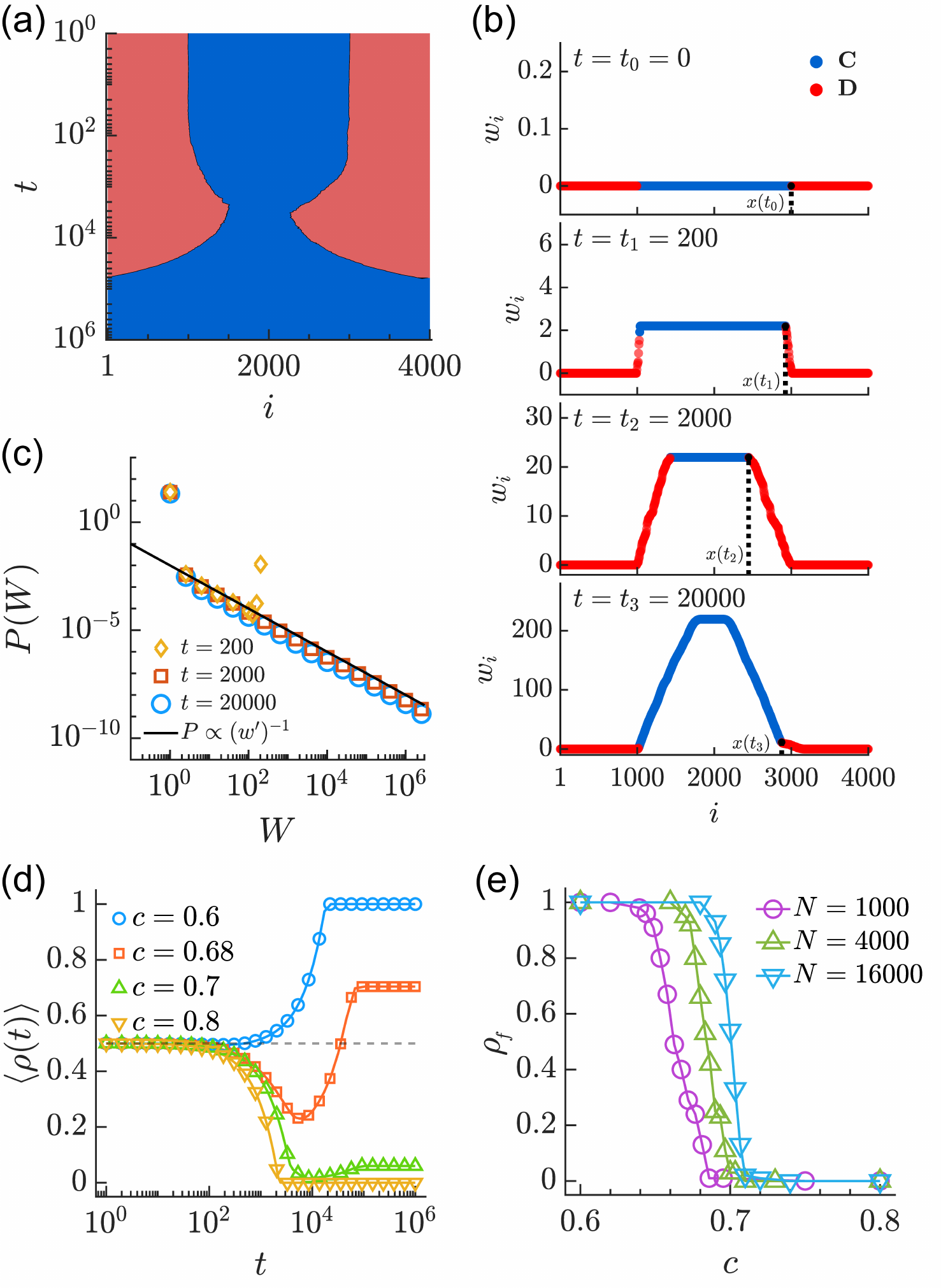}
\caption{
Evolution of strategy and wealth. 
(a) Evolution of  the strategy configuration
$\{s_i(t)|1\leq i\leq N=4000, 1\leq t\leq T=10^6\}$
in a single run of simulation under periodic boundary condition.
Blue (red) color represents cooperation (defection).
Initially $s_i={\rm C}$ for $i\in \left[\frac{N}{4}+1, \frac{3N}{4} \right]$
and $s_i={\rm D}$ otherwise.  
(b) Log-wealth profiles $\{w_i(t)=\log_{10}W_i \mid 1\le i\le N\}$
at times $t_0,t_1,t_2,t_3$, from a single run.
$x(t_i)$ marks the boundary between the central $C$ cluster
and the right $D$ cluster at time $t_i$.
(c) Wealth distribution at different times.
(d) Time-evolution of the density of cooperators $\rho(t)$. 
(e) Frequency of fixation to cooperation versus cost-benefit ratio $c$. 
All results are obtained at $\varepsilon=0.01$, $\beta=100$, $c=0.68$,
and $N=4000$ unless stated otherwise and the results in (c-e) are averaged over 100 simulation runs.
 }
\label{fig:simulation}
\end{figure}

In Monte Carlo simulations [Fig.~\ref{fig:simulation}],
each run eventually fixates to  either all-C ($s_i={\rm C}$ for all $i$)
or all-D ($s_i={\rm D}$ for all $i$).
Remarkably, fixation to cooperation can occur even at $c>c_\crit^\nowealth=1/2$,
where the cooperator domain undergoes an early contraction, followed by regrowth,
and ultimately expands to occupy the entire system [Fig.~\ref{fig:simulation}(a)].
Simultaneously, individuals' wealth---initially uniform---becomes heterogeneous over
time: between the persistent cooperators, whose wealth grows exponentially with time,
and the persistent defectors, whose wealth remains at the initial value,
lie individuals who have switched strategies,
with their wealth decaying exponentially with distance from the persistent cooperator region
[Fig.~\ref{fig:simulation}(b)].
Consistent with this exponential wealth profile, the wealth distribution,
initially a delta function $P(W) = \delta(W-1)$,
evolves towards a power-law $P(W)\sim W^{-1}$ [Fig.~\ref{fig:simulation}(c)].

The ensemble-averaged cooperator density,
$\langle\rho(t)\rangle=\langle \frac{1}{N}\sum_{i=1}^{N}\delta_{s_i(t), {\rm C}}\rangle$
initially decreases and then regrows for moderately large $c>1/2$ [Fig.~\ref{fig:simulation}(d)].
The final value $\rho_\fixation=\langle \lim_{t\to\infty} \rho(t)\rangle$,
corresponding to the fixation frequency of cooperation,
remains close to $1$ over a broad range of $c$.
The threshold $c_\crit$,
at which $\rho_\fixation=1/2$,
is greater than $c_\crit^\nowealth=1/2$ [Fig.~\ref{fig:simulation}(e)] and may vary with $N$ unlike  $c_\crit^\nowealth$.

\section{Results}
\label{sec:results}

\subsection{Wealth gradient dictates fixation}
\label{sec:gradient}

Fixation to cooperation or defection is governed by 
 the motion of the C-D boundary separating the cooperator(C)-domain ($i\leq {\boundary}$) and the defector(D)-domain ($i\geq {\boundary}+1$), which behaves as a stochastic interface coupled to accumulated wealth. Hereafter $x$ denotes the position of the C-D boundary,
representing the location of the rightmost cooperator.
It can move right, expanding the C-domain, when the leftmost defector at $x+1$ flips to cooperation
or move left, expanding the D-domain, when the rightmost cooperator at $x$ flips to defection. 
Such shifts of the boundary occur with probabilities $P_\Coop$ and $P_\Defect$, respectively, obtained from Eq.~\eqref{eqn:copy_probability} as
\begin{equation}\label{eqn:P_D_P_C}
\begin{split}
  & P_\Coop  = P_{{\boundary}}^{{{\boundary}}+1}=\frac{1}{1+e^{\beta \varepsilon\{(1+r_{{\boundary}}) c-r_{{\boundary}}\}}},\\
  & P_\Defect = P_{{\boundary}+1}^{{\boundary}}= \frac{1}{1+e^{\beta \varepsilon\{(1+r_{{{\boundary}}-1})(1-c)-r^{-1}_{{\boundary}}\}}}.
\end{split}
\end{equation}
Here we assumed $W_{i}\geq W_{i+1}$ for all $i$ near the C-D boundary [Fig.~\ref{fig:simulation}(b)]
to obtain the payoffs
$f_{{{\boundary}}+1}^{{{\boundary}}}=\varepsilon W_{{{\boundary}}+1}, f_{{{\boundary}}-1}^{{{\boundary}}}
= \varepsilon (1-c) (W_{{{\boundary}}-1}+W_{{\boundary}}), f_{{{\boundary}}}^{{{\boundary}}+1}
= (1-c) W_{{\boundary}} - c W_{{{\boundary}}+1}$, and $f_{{{\boundary}}+2}^{{{\boundary}}+1} = 0$
and we introduced $r_i \equiv \frac{W_i}{W_{i+1}}$ as a measure of the steepness of the wealth profile,
and hence of local wealth inequality.
Hereafter we refer to $r_i$ as the wealth gradient.
The C-D boundary is more likely to move right (left),
leading to fixation to cooperation (defection),
if $P_\Coop>P_\Defect$ ($P_\Coop<P_\Defect$) or equivalently if the boundary wealth gradients
$r_\boundary$ and $r_{\boundary-1}$ (assumed equal) lie above (below)
the threshold
$r_\threshold \equiv {1\over 2(1-c)}$.
We will focus on the nontrivial regime  $c>1/2$;
for $c<1/2$, since $r_\threshold<1$ and $P_\Coop>P_\Defect$, fixation to cooperation readily follows.

\begin{figure}[t]
\ff{1.}{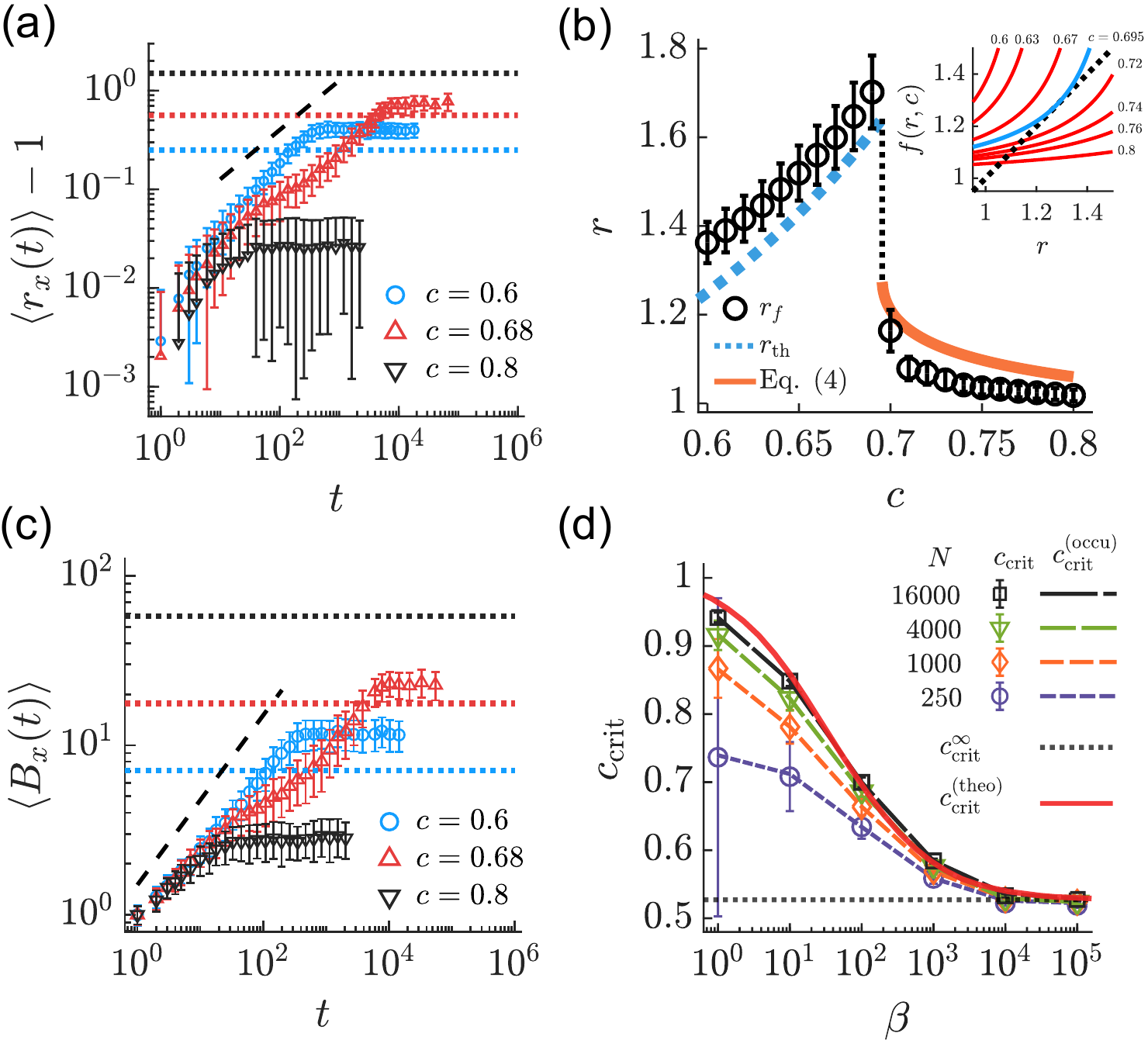}
\centering
\caption{
Wealth gradient and phase diagram in one dimension.
(a) Excess wealth gradient $r_x(t)-1$ at the domain boundary for different cost-benefit ratios $c$.
Dotted lines represent $r_\threshold-1={1\over 2(1-c)}-1$ and the dashed line has a slope $1/2$.
(b) Saturated wealth gradient $r_\fixation$ versus $c$ from simulations (open)
and analytic prediction $r_\fixation = r_\threshold={1\over 2(1-c)}$ for $c<c_\crit^\theoretical$ (dotted)
and the solution of Eq.~\eqref{eq:self} for $c>c_\crit^\theoretical$ (solid)
with $c_\crit^\theoretical \approx 0.695$.
Inset: Plots of $y=r$ and $y=f(r,c)$ with $f(r,c)$ given by the r.h.s. of Eq.~\eqref{eq:self}
for different values of $c$.
(c) Boundary occupation time $B_\boundary(t)$.
Dotted line represent $B_\threshold$ and the dashed line has slope $1/2$. 
(d) Critical ratios $c_\crit$ versus inverse temperature $\beta$,
dividing the cooperation regime $c<c_\crit(\beta)$
and the defection regime $c>c_\crit(\beta)$ for $\varepsilon=0.01$
and different system sizes $N$.
Shown are $c_\crit$ (points) and $c_\crit^\occupation$ (dashed) from simulations,
$c_\crit^\theoretical$ (solid) from Eq.~\eqref{eq:self}, and $c_\crit^{\infty}$ in the
$\beta\to\infty$ limit from Eq.~\eqref{eqn:beta_inf}.
All results are averaged over $100$ simulation runs and $\beta=100,\varepsilon=0.01$
and $N=4000$ in (a-c). 
}
\label{fig:boundary}
\end{figure}

The wealth gradient $r_\boundary$ increases with time and saturates [Fig.~\ref{fig:boundary}(a)].
For $c>c_\crit$, the saturation value $r_\fixation=\lim_{t\to\infty} \langle r_{{\boundary}}(t)\rangle$
is smaller than $r_\threshold$, indicating that the C-D boundary drifts left and the system typically
fixates to all-D.
In contrast, for $1/2<c<c_\crit$,
the boundary wealth gradient increases sufficiently to exceed
the threshold, when the C-D boundary reverses direction from left to right and drifts rightward,
typically leading the system to fixation in the all-C state. [Fig.~\ref{fig:simulation}(a,d)].

To determine the conditions under which the wealth gradient exceeds or remains below the threshold,
we begin by noting from Fig.~\ref{fig:simulation}(b) an approximately spatially uniform gradient
$r$ between the persistent cooperators far to the left of the C-D boundary,
whose wealth grows exponentially as
$W= [1+8\varepsilon(1-c)]^t$, and persistent defectors far
to the right, with wealth fixed at $W=1$.
Hence $r\simeq \pas{1+8\varepsilon(1-c)}^{t\over{\Delta{\boundary}}(t)}$,
where $\Delta {\boundary}(t)$ denotes the number of sites visited by the C-D boundary up to time $t$,
i.e., sites where strategies have switched at least once.
Suppose that the boundary keeps drifting left with a constant velocity
$v=P_\Coop-P_\Defect<0$~\footnote{Contributions to $v$ from higher order terms in $P_\Defect$
and $P_\Coop$ are neglected.},
where  $r_\boundary=r_{\boundary-1}=r<r_\threshold$.
Then $\Delta \boundary(t)\simeq |v|t$ in the long time limit, and we have 
\begin{equation}
r = \pas{1+8\varepsilon(1-c)}^{1\over P_\Defect-P_\Coop} .
\label{eq:self}
\end{equation}
Since $P_\Defect-P_\Coop$ depends on $r$, Eq.~\eqref{eq:self} is self consistent
and determines $r$ when a solution exists.
For given $\beta$ and $\varepsilon$, there is a critical ratio $c_\crit^\theoretical$:
for $c>c_\crit^\theoretical$ two solutions exist, one stable and one unstable,
at $c=c_\crit^\theoretical$ they merge into a semistable solution $r_\crit^\theoretical$,
and for $c<c_\crit^\theoretical$ no solution exists,
which corresponds to a saddle-node bifurcation~\cite{strogatz:2000}.
For $\beta=100$ and $\varepsilon=0.01$, $c_\crit^\theoretical=0.695436\ldots$,
as shown in Fig.~\ref{fig:boundary}(b).
The critical ratio $c_\crit^\infty$ in the $\beta\to \infty$ limit is derived in Appendix~\ref{sec:betainfinite} 
and further properties of $c_\crit^\theoretical$ for finite $\beta$ are analyzed in Appendix~\ref{sec:betafinite}.
The stable solution approximates well the saturated wealth gradient  $r_\fixation$ from simulations
for $c>c_\crit\approx c_\crit^\theoretical$ [Fig.~\ref{fig:boundary}(b)].
The loss of stability at $c_\crit^\theoretical$ implies that the C-D boundary can no longer
drift left persistently, and that the wealth gradient cannot remain below $r_\threshold$.

\subsection{Boundary stalling and discontinuous wealth-gradient transition}
\label{sec:boundary}

We find that $r_\fixation$ slightly
exceeds $r_\threshold$ for $c<c_\crit$, displaying an abrupt jump of size
$\Delta r \approx r_\threshold - r_\crit^\theoretical$ at the critical ratio [Fig.~\ref{fig:boundary}(b)].
To understand how such a large wealth gradient arises for $c<c_\crit$ with a discontinuous
transition at the critical ratio, we adopt a more refined approximation for individuals' wealth
than the one leading to Eq.~\eqref{eq:self}.
An individual's wealth $W_i(t)$ grows multiplicatively when it cooperates,
neglecting the rare events of gaining an exploitation payoff by defecting against
a cooperating neighbor or incurring the opposite sucker's payoff.
Thus we approximate
$W_i(t) \simeq \pas{1 + 8\varepsilon(1-c)}^{C_i(t)}$,
where $C_i(t) \equiv \sum_{t'=0}^{t-1} \delta_{s_i(t'),{\rm C}}$ counts the number of cooperative
time steps up to time $t$.
This reproduces well the true wealth profile 
and yields the local wealth gradient given by 
\begin{equation}\label{eqn:r(t)}
    r_i(t) \approx \left[ {1 + 8\varepsilon(1-c)} \right]^{B_{i}(t)},
\end{equation}
where  $B_i(t) \equiv  C_i(t) - C_{i+1}(t) =\sum_{t'=0}^{t-1} \delta_{i,{{\boundary}}(t')}$
denotes the occupation (visit) time of the C-D boundary at site  $i$ up to time $t$. This approximation uncovers the link between the wealth gradient and the domain boundary's motion.

Let us first consider the limit $\beta\to\infty$.
The velocity of the C-D boundary, $v=P_\Coop-P_\Defect$, is $-1$
if $r_{{\boundary}}<r_\Defect$, $v=0$ if $r_\Defect<r_{{\boundary}}<r_\Coop$,
and $v=1$ if $r_{{\boundary}}>r_\Coop$ with two thresholds $r_\Defect$ and $r_\Coop$ given in End Matter.
When $c$ is slightly below $c_\crit^\infty$, the C-D boundary initially moves left ($v=-1$)
since $r_i(0)=1<r_\Defect$ everywhere, but then immediately stalls as the boundary wealth gradient $r_x(1)$ at $t=1$ exceeds $r_\Defect$ (yet remains below $r_\Coop$),  resulting in $v=0$. With the boundary ${\boundary}$ fixed,  the occupation time $B_{{\boundary}}(t)$  grows linearly with time and $r_{{\boundary}}(t)$ increases exponentially by Eq.~\eqref{eqn:r(t)}, quickly surpassing $r_\Coop$ and driving the boundary rightward ($v=1$) toward fixation to cooperation. 

Similarly, when $\beta$ is finite and $c$ is below $c_\crit$, the C-D boundary initially moves left but slows down (Appendix~\ref{sec:velocity}) due to the increase of the boundary wealth gradient and then effectively stalls. This  increases explosively the wealth gradient. The occupation time at the boundary $B_{{\boundary}}(t)$ is found to grow sublinearly with time as $B_{{\boundary}}(t) \sim t^{\nu}$ with  
$\nu\simeq 0.44$ for $c=0.6$ and $\nu\simeq 0.48$ for $c=0.68$
at $\beta=100$ and $\varepsilon=0.01$ [Fig.~\ref{fig:boundary}(c)]. This implies that the  boundary motion scales like a random walk, visiting $S(t)\sim t^{1/2}$ distinct sites in time $t$, with each site visited on average ${t\over S(t)}\sim t^{1/2}$ times. Consequently, the boundary wealth gradient grows in a stretched-exponential manner and surpasses the threshold $r_\threshold$ [Fig.~\ref{fig:boundary}(a)], leading to $P_\Coop > P_\Defect$ and driving the boundary rightward.
This process can repeat all the way until the C-domain expands to occupy the entire system
by satisfying $r_i>r_\threshold$ at sites visited by the C-D boundary.
This mechanism explains how the wealth gradient grows to exceed slightly the threshold  $r_\fixation\approx r_\threshold$ for $c\leq c_\crit$ as well as the origin of the gap $\Delta r_\crit \approx r_\threshold - r_\crit^\theoretical$ at $c_\crit$ [Fig.~\ref{fig:boundary} (a,b)]. Note that for $c>c_\crit$, the boundary persistently moves left with almost constant velocity (Appendix~\ref{sec:velocity}) and the boundary occupation time saturates fast at a small value [Fig.~\ref{fig:boundary}(c)].

\subsection{Influence of temperature and system size}
\label{sec:temperature}

This interplay between the boundary motion and the wealth gradient can be influenced by temperature.
If temperature increases ($\beta$ decreases), the flip probabilities
of Eq.~\eqref{eqn:P_D_P_C}
approach $1/2$ and
converge toward each other,
slowing the boundary motion while enhancing its fluctuations.
The resulting increase of the boundary occupation time can shift the critical ratio $c_\crit$ to larger values, thereby further promoting cooperation.
In agreement with this prediction,
simulations show that the critical ratio $c_\crit$ increases with
decreasing $\beta$ [Fig.~\ref{fig:boundary}(d); see also Appendix~\ref{sec:varepsilon} for results at different $\varepsilon$].
Moreover, to corroborate our analysis, we also measured $c_\crit^\occupation$
at which $\rho_\fixation^\occupation \equiv \ave{\Theta\pas{B_\fixation - B_\threshold}}$
is equal to $1/2$  with $B_\fixation$,
the saturated boundary occupation time and
$B_\threshold \equiv {\ln r_\threshold\over \ln\pas{1+8\varepsilon (1-c)}}$ from Eq.~\eqref{eqn:r(t)},
and $c_\crit^\theoretical$ from the bifurcation analysis of Eq.~\eqref{eq:self} (Appendix~\ref{sec:betafinite}),
which all agree well with $c_\crit$ measured directly from simulations.

In conventional scenarios, high temperature suppress cooperation by increasing encounters between cooperators and defectors, thereby hindering the assortment of cooperators and favoring defectors~\cite{szabo2007evolutionary,li2021limited}. 
In our model on a one-dimensional lattice, this negative effect is constrained  but instead finite-temperature fluctuations enhance cooperation by increasing the boundary occupation time, and thereby elevating the wealth of cooperators relative to defectors. 

Both $c_\crit$ and $c_\crit^\occupation$ reveal finite-size effects in Fig.~\ref{fig:boundary}(d) indicating that large system size helps the emergence of cooperation. This can be understood by considering that for $c>1/2$, a small system can quickly fixate to all-D before developing a sufficiently steep wealth gradient. 

\begin{figure}[t]
\centering
\ff{1.}{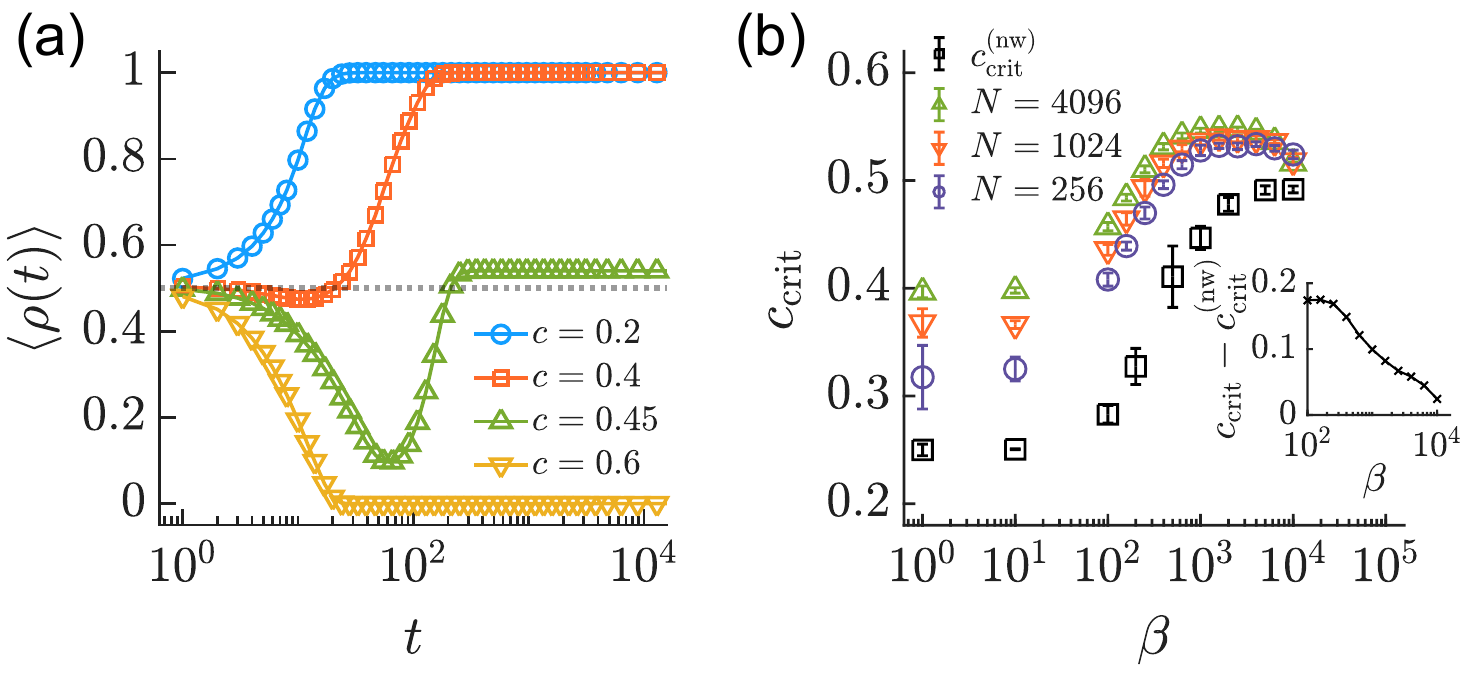}
\caption{Evolution of cooperator density and phase diagram in two dimension. 
(a)
  Time-evolution of the cooperator density $\rho(t)$ for different cost-benefit ratios $c$
  from simulations on the two-dimensional lattice of size
  $N=L\times L$ with $N=4096 (L=64)$,
  $\beta=100$, and $\varepsilon=0.01$. 
(b) 
  Critical ratios $c_\crit$ versus inverse temperature  $\beta$ for different system sizes with
  $\varepsilon=0.01$. The critical ratios without wealth-scaled interactions $c_\crit^{\nowealth}$
  for $N=4096$ is also shown (square). 
  The inset shows the difference between $c_{\rm crit}$ and $c_{\rm crit}^{\rm (nw)}$ as a function of $\beta$.
}
    \label{fig:2D}
\end{figure}

\subsection{Optimality in two dimensions}
\label{sec:2d}

Our findings in one dimensions are largely reproduced
in two dimensions. Without wealth-scaled interactions,
the critical ratio is given by $c_\crit^\nowealth = 1/4$
in the high-temperature limit $\beta \to 0$~\cite{ohtsuki_2006}.
In our model with wealth-scaled interactions, the density of cooperators exhibits regrowth after initial decrease even for  $c=0.45$ [Fig.~\ref{fig:2D}(a)] as in one dimensions. Moreover, the critical ratio $c_\crit$  is much larger than that without wealth-scaled interactions [Fig.~\ref{fig:2D}(b)]. The increase of the critical ratio $c_\crit - c_\crit^\nowealth$ due to wealth-scaled interactions is stronger for smaller $\beta$, indicating that cooperation promotion under wealth-scaled interactions is stronger at higher temperature as in one dimensions.  In two dimensions, however, C- and D-domains can be fragmented into multiple domains (Appendix~\ref{sec:2d}) in contrast to one dimension, so that defection can be favored at high temperatures. Given these competing roles of temperature, we expect the existence of an optimal temperature at which $c_\crit$ is maximized. This is indeed the case as shown in Fig.~\ref{fig:2D}(b); $c_\crit$ is maximized at $\beta\approx 3000$ for $\varepsilon=0.01$. 

\section{Summary and discussion}
\label{sec:summary}

We have elucidated a mechanism by which wealth accumulated from individuals' past strategic choices can promote cooperation.  By scaling interaction payoffs by the participants' minimum wealth, frequent cooperators accumulate greater wealth than frequent defectors, offsetting the immediate advantage of defection.  We have shown that when the boundary between cooperator  and defector domains moves slows and effectively stalls, the wealth gradient grows rapidly and saturates, driving the boundary back and expanding the cooperator domain in one dimension. The condition for such effective stalling has been derived analytically, revealing a discontinuous transition of the saturated wealth gradient at the critical ratio.  

The interplay between the wealth gradient and cluster-level interactions uncovered in this study represents a novel pathway to the emergence of cooperation in societies. 
Moreover, wealth heterogeneity and stochastic boundary dynamics fundamentally alter the role of fluctuations,
turning them from a disruptive factor into a driver of cooperation.
To consolidate this framework,  the analysis can be extended to alternative settings such as birth-death updating  and higher spatial dimensions. Extending the analytical theory to higher dimensions and heterogeneous structures is an important direction for future research. It is also important to examine the robustness of our results under different payoff scalings, for instance, where payoffs are constrained by a participant's own wealth or  that of its partner. 
Another direction is to  generalize pairwise  to group interactions  and consider higher-order interaction networks, thereby deepening our understanding of the wealth gradient and boundary dynamics.

\begin{acknowledgements}
This work was supported in part by the National Research Foundation of Korea (NRF) grant funded by the Korea government (MSIT) [No. RS-2024-00359230(H.-C.J.)] and by KIAS Individual Grants [Nos.~CG084502 (H.G.L.) and CG079902 (D.-S.L.)] at Korea Institute for Advanced Study. We thank the Center for Advanced Computation in KIAS for providing the computing resources.
\end{acknowledgements}




\appendix
\numberwithin{equation}{section} 
\numberwithin{figure}{section}

\section{Critical ratio in the $\beta\to\infty$ limit}
\label{sec:betainfinite}

In the limit $\beta\to\infty$, flips are deterministic,
$P_\Coop=0$ for $r<r_\Coop$ and $P_\Coop=1$ for $r>r_\Coop$,
and $P_\Defect=1$ for $r<r_\Defect$ and $P_\Defect=0$ for $r>r_\Defect$,
where $r_\Coop={c\over 1-c}$ and $r_\Defect={1\over 2}\left(\sqrt{5-c\over 1-c}-1\right)$.
Both increase with $c$ and meet at $r_\Coop=r_\Defect=1$ when $c=1/2$,
hence $0\le r_\Coop<r_\Defect<1$ for $0\le c<1/2$ and $1<r_\Defect<r_\Coop$ for $1/2<c<1$.

Recalling that the wealth gradient $r_i = W_i/W_{i+1}$ is not smaller than $1$ around the C-D boundary,
we find that for $c<1/2$,  both thresholds are smaller than $1$ and thus $P_\Coop=1$ and $P_\Defect=0$,
resulting in the C-D boundary moving right and fixation to cooperation.
For $c>1/2$, those thresholds become larger than $1$ and moreover,
$1<r_\Defect < r_\Coop$, implying that the boundary may move right or left, or stall, i.e.,
its typical (average) velocity $v=  P_\Coop-P_\Defect$  is given by 
\begin{equation}
v = \left\{
\begin{array}{ll}
-1 & \ {\rm for} \ 1\leq r<r_\Defect,\\
0 & \ {\rm for} \ r_\Defect <r<r_\Coop,\\
1 & \ {\rm for} \ r>r_\Coop.
\end{array}
\right.
\label{eq:vinf}
\end{equation}

When the C-D boundary moves left, its velocity is necessarily $v=-1$ from Eq.~\eqref{eq:vinf}, and thus the wealth gradient is given by $r=1+8\varepsilon(1-c)$ from Eq. \eqref{eq:self}. To be self-consistent, the gradient should be smaller than $r_\Defect$ according to Eq.~\eqref{eq:vinf}, which holds when $c>c_\crit^\infty$ with $c_\crit^\infty$ satisfying
\begin{equation}
  \label{eqn:beta_inf}
  r_\Defect(c_\crit^{\infty})
  \equiv {1\over 2} \left(\sqrt{5-c_\crit^\infty\over 1- c_\crit^\infty}-1\right)
  = 1+8\varepsilon(1-c_\crit^{\infty}).
\end{equation}

The domain boundary velocity takes the values $v=1, -1,$ or $0$, in the limit $\beta\to\infty$,
which simplifies the dynamics and helps us understand that when $c$ is smaller than
$c_\crit^\infty$, the system fixates to cooperation in the $\beta\to\infty$ limit.
Suppose that $c<c_\crit^\infty$ or equivalently, $r_\Defect<1+8\varepsilon (1-c)$.
 At $t=0$, the wealth gradient is $r=1$ everywhere, so  the C-D boundary initially moves one step to the left ($v=-1$), giving $r_{{\boundary}}(t=1) \approx 1+8\epsilon (1-c)$, consistent with Eq.~\eqref{eqn:r(t)} using $B_{{\boundary}}(t=1)=1$. (i) If $r_{{\boundary}}(t=1)$  exceeds $r_\Coop$, then by Eq.~\eqref{eq:vinf}, the domain boundary reverses direction at $t=1$ and moves right ($v=1$) for $t\geq 1$. At the boundary site visited at each time step, $B_{{\boundary}}=1$ due to $v=1$ and thus the wealth gradient is given by $r_{{\boundary}}=1+8\varepsilon(1-c)$, larger than $r_\Coop$. (ii) If $r_\Defect<r_{{\boundary}}(t=1)<r_\Coop$,  then the domain boundary stalls at ${{\boundary}}(t=1)$, increasing the occupation time $B_{{\boundary}}$. Consequently, $r_{{{\boundary}}}$ grows and eventually exceeds $r_\Coop$. Then the boundary is pushed one step right since $P_\Coop=1$ and $P_\Defect=0$ (equivalently $v=1$). At the new position, the C-D boundary can stall again, increasing $B_{{\boundary}}$ until $r_{{\boundary}}$ exceeds $r_\Coop$, after which it moves right. This stall-and-move process repeats, eventually driving the system to fixation at all-C.

\section{Critical ratio for finite $\beta$}
\label{sec:betafinite}

Suppose that the C-D domain boundary keeps drifting left, without changing the direction,
and the wealth gradient saturates quickly at a value $r$ far below  $r_\threshold$,
leading to fixating to defection.
In this case, the velocity of the domain boundary 
\begin{align}
  v(&r;c,\beta,\varepsilon) \equiv P_\Coop - P_\Defect
  \nonumber\\
  &= \frac{1}{1+e^{\beta \varepsilon\{(1+r) c-r\}}}-\frac{1}{1+e^{\beta \varepsilon\{(1+r)(1-c)-r^{-1}\}}}
\end{align}
is negative and nearly constant against time.
Then the wealth gradient $r$ satisfies Eq.~(4) of the main text or equivalently 
\begin{equation}
  r = f(r; c, \beta,\varepsilon) \ {\rm with}
  \ f(r;c,\beta,\varepsilon) \equiv [1+8\epsilon (1-c)]^{1\over |v(r;c,\beta\varepsilon)
|}
\label{eq:rfrcbe}
\end{equation}
for given parameters $c,\beta$, and $\varepsilon$.
Note that Eq.~(4) or Eq.~\eqref{eq:rfrcbe} is a self-consistent equation for the
wealth gradient $r$ under the assumption of fixation to defection.
The function $f(r)$ monotonically increases with $r$ from
$f(1)=[1+8\varepsilon (1-c)]^{1\over |v(1)|}$ and diverges at
$r=r_\threshold$ (for $\beta$ finite) or $r_\Defect$ (for $\beta$ infinite).
The stable fixed point satisfying $r=f(r)$ with ${\partial f\over \partial r}<1$ gives
the wealth gradient.
The curves $y=f(r)$ and $y=r$ meet at two points for $c>c_\crit^\theoretical$, tangentially for
$c=c_\crit^\theoretical$, and do not intersect for $1/2<c<c_\crit^\theoretical$ as shown in the
inset of Fig. 2(b),
suggesting that a stable fixed point of
Eq.~(4) exists only for $c>c_\crit^\theoretical$.
The critical point, i.e.,
the threshold cost-benefit ratio $c_\crit^\theoretical$,
and the wealth gradient $r_\crit^\theoretical$ at the critical point are obtained by solving
Eq.~\eqref{eq:rfrcbe} (or Eq. (4) in the main text) and 
\begin{equation}
  {\partial f\over \partial r}
  = r \ln [1+8\varepsilon (1-c)] {1\over v^2} {\partial v\over \partial r}=1
\label{eq:df}
\end{equation}
simultaneously, as illustrated in Fig.~2(b).
Although closed-form solutions are not available in general,
their asymptotic behaviors can be derived for very large or small $\beta$. 

\begin{figure}
\ff{0.8}{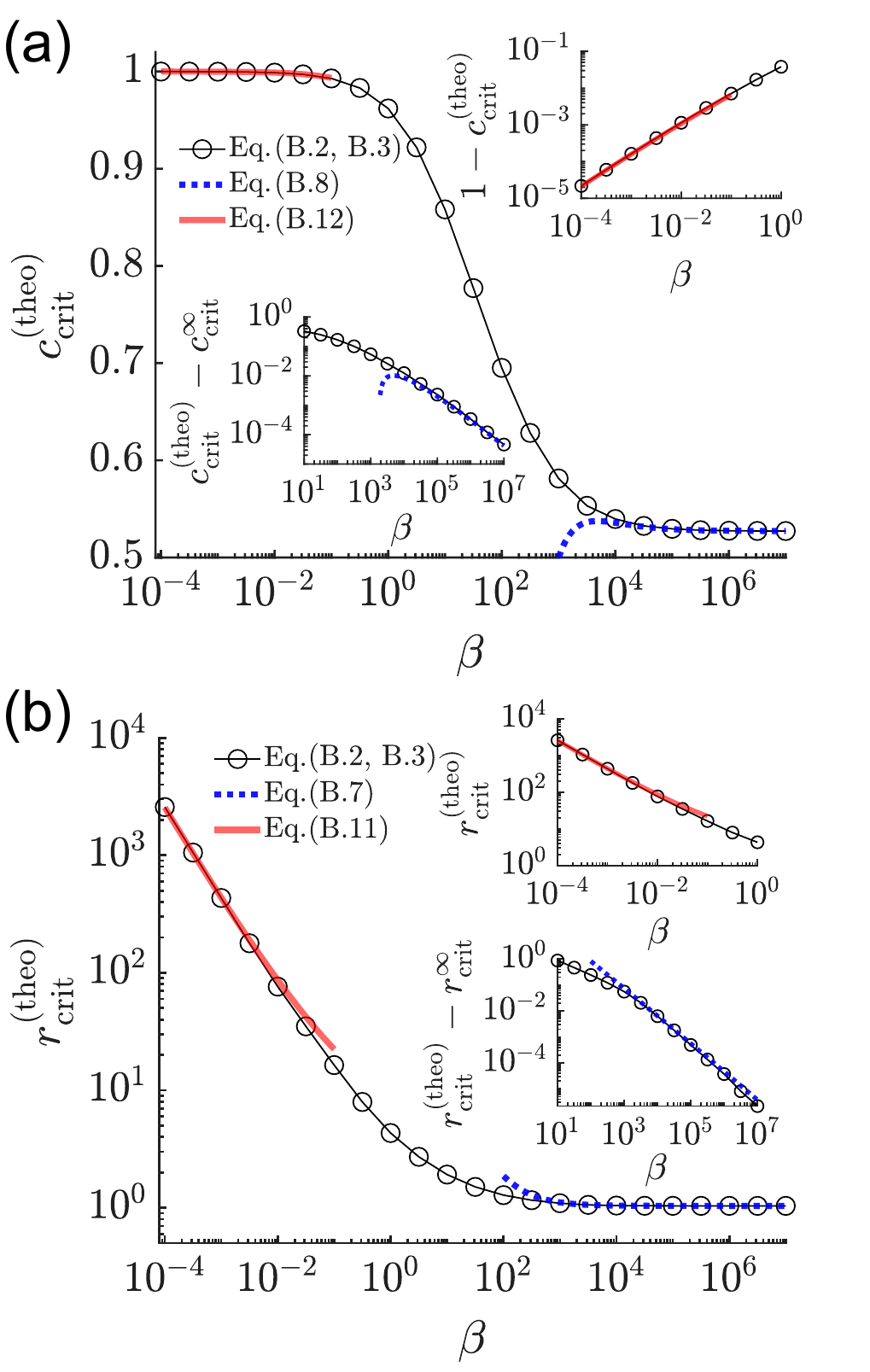}
\caption{
  Temperature-dependent critical points and their asymptotic behaviors for $\varepsilon=0.01$.
  (a) The critical point $c_\crit^\theoretical$ as a function of inverse temperature $\beta$ obtained
  by solving numerically Eqs.~\eqref{eq:rfrcbe} and \eqref{eq:df} (points) and its asymptotic
  behaviors for $\beta$ large and small given by Eq.~\eqref{eq:ccrit_large} (dashed)
  and Eq.~\eqref{eq:ccrit_small} (solid), respectively.
  Inset: Plots of $1-c_\crit^\theoretical$ versus $\beta$ from numerical solutions (points)
  and Eq.~\eqref{eq:ccrit_small} (line)
  (b) The wealth gradient $r_\crit^\theoretical$ in the limit $c\downarrow  {c_\crit^\theoretical}$ versus
  $\beta$ from numerical solutions (points) and its asymptotic behaviors given
  by Eq.~\eqref{eq:rcrit_large} (dashed) and Eq.~\eqref{eq:rcrit_small} (solid).
}
\label{fig:critical}
\end{figure}

\subsection{Solutions in the low temperature limit ($\beta \gg 1$)}

We have derived the critical ratio in the $\beta\to\infty$ limit in Eq.~\eqref{eqn:beta_inf} in Appendix~\ref{sec:betainfinite}. Also the wealth gradient at the critical point $r_\crit^\infty$,
defined in the limit 
$c \downarrow c_\crit^\infty$(approaching $c_\crit^\infty$ from above),
is given by
\begin{equation}
r_\crit^\infty = r_\Defect (c_\crit^\infty) = 1+8\varepsilon (1-c_\crit^\infty).
\label{eq:rcrit_infty}
\end{equation}
If $\varepsilon\ll 1$, we have approximately
$c_\crit^\infty \simeq {1\over 2} + 3\varepsilon$ and
$r_\crit^\infty = r_\Defect = 1+4\varepsilon$,
yielding e.g., $c_\crit^\infty \simeq 0.53$
and $r_\crit^\infty \simeq 1.04$ at $\varepsilon=0.01$.

To explore the leading correction to $c_\crit^\infty$ and
$r_\crit^\infty$ for $\beta$ large but finite,
we consider
$r = r_\crit^\infty + \delta r$
and
$c = c_\crit^\infty + \delta c$.
When $\delta c$ and $\delta r$ are small, we have $P_C\approx 0$ and
the boundary velocity given approximately by $v\simeq -P_D$.
Up to the leading order in $\delta r$ and $\delta c$, we obtain
\begin{align}
  v
  &\simeq -1 + e^{\beta \varepsilon \{(1-c_\crit^\infty +(r_\crit^\infty)^{-2})\delta r
    - (1+r_\crit^\infty)\delta c\}},\nonumber\\
  {\partial v\over \partial r}
  &\simeq  \beta \varepsilon (1-c_\crit^\infty + (r_\crit^\infty)^{-2})) \nonumber\\
   & \times e^{\beta \varepsilon \{(1-c_\crit^\infty +(r_\crit^\infty)^{-2})\delta r - (1+r_\crit^\infty)\delta c\}},
\label{eq:largeapprox}
\end{align}
where we assumed $\beta\varepsilon\, \delta c\gg 1$.
We also see from Eq.~\eqref{eq:rfrcbe} that
\begin{equation}
  \delta r
  \simeq - 8\varepsilon \delta c + r_\crit^\infty
  \ln \pas{r_\crit^\infty  e^{\beta \varepsilon \{(1-c_\crit^\infty +(r_\crit^\infty)^{-2})\delta r - (1+r_\crit^\infty)\delta c\}}}.
\end{equation}
Using these results and the assumption $\delta r\ll \delta c$ in Eq.~\eqref{eq:df}, we obtain
\begin{equation}
r_\crit^\theoretical \simeq r_\crit^\infty - 8\varepsilon \delta c + {1\over \beta \varepsilon (1-c_\crit^\infty + (r_\crit^\infty)^{-2})},
\label{eq:rcrit_large}
\end{equation}
and
\begin{equation}
c_\crit^\theoretical  \simeq c_\crit^\infty +{\ln [r_\crit^\infty(1-c_\crit^\infty + r_\Defect^{-2}) \beta \varepsilon \ln r_\crit^\infty] \over (1+r_\crit^\infty) \beta \varepsilon}.
\label{eq:ccrit_large}
\end{equation}
These asymptotic behaviors are in good agreement with the true solutions as shown in Fig.~\ref{fig:critical}.

\subsection{Solutions in the high temperature limit ($\beta\ll 1$)}

When $\beta$ is small, the boundary's velocity $v$ behaves as 
\begin{align}
  v&\simeq - {\beta \varepsilon \over 4} \{2c -1 - 2(1-c) r + r^{-1}\},\nonumber\\
  {\partial v\over \partial r}&\simeq {\beta\varepsilon \over 4} \{2(1-c) +r^{-2}\},
\end{align}
and the solution $r$ of Eq.~\eqref{eq:rfrcbe} is expected to be large.
Using these results in Eq.~\eqref{eq:df}, we obtain
\begin{align}
  r &\ln [1+8\varepsilon (1-c)] {1\over v^2} {\partial v\over \partial r}
  \simeq \nonumber\\
  &r \ln [1+8\varepsilon (1-c)] {4\over \beta\varepsilon}
         {2(1-c)+r^{-2} \over [2c-1 -2(1-c)r +r^{-1}]^2}=1.
\end{align}
Considering that $r$ and $1/(\beta\varepsilon)$ are large,
we find that $1-c$ must be small.
Let $\delta c = 1-c$ and assume $\delta c\ll 1$.
Then we obtain the approximations
$v\simeq -\frac{\beta\varepsilon}{4},(1-2r\delta c)$,
$\frac{\partial v}{\partial r} \simeq \frac{\beta\varepsilon}{2},\delta c$,
and $r\simeq \exp \big\{\frac{32\delta c}{\beta(1-2r\delta c)}\big\}$.
Define $z=\frac{32\delta c}{\beta(1-2r\delta c)}$.
Then $r\simeq e^{z}$.
Substituting these into Eq.~\eqref{eq:df} gives
$e^{z}z^{2}\simeq \frac{16}{\beta}$
and solving this yields
\begin{equation}
  z = \ln r_\crit^\theoretical \simeq \ln {16\over \beta}
  - 2 \ln \ln {16\over \beta} + 4 {\ln \ln {16\over \beta} \over \ln {16\over \beta}},
\label{eq:rcrit_small}
\end{equation}
and $\delta c \simeq z {\beta \over 32}(1-2 r \delta c)$ leading to 
\begin{equation}
  c_\crit^\theoretical \simeq 1 - {\beta \over 32}
  \left[ \ln {16\over \beta}  - 2 \ln \ln {16\over \beta} -1
    + 4 {\ln \ln {16\over \beta} \over \ln {16\over \beta}}
    \right].
\label{eq:ccrit_small}
\end{equation}

\section{Velocity of the C-D boundary}
\label{sec:velocity}

\begin{figure}
\ff{.7}{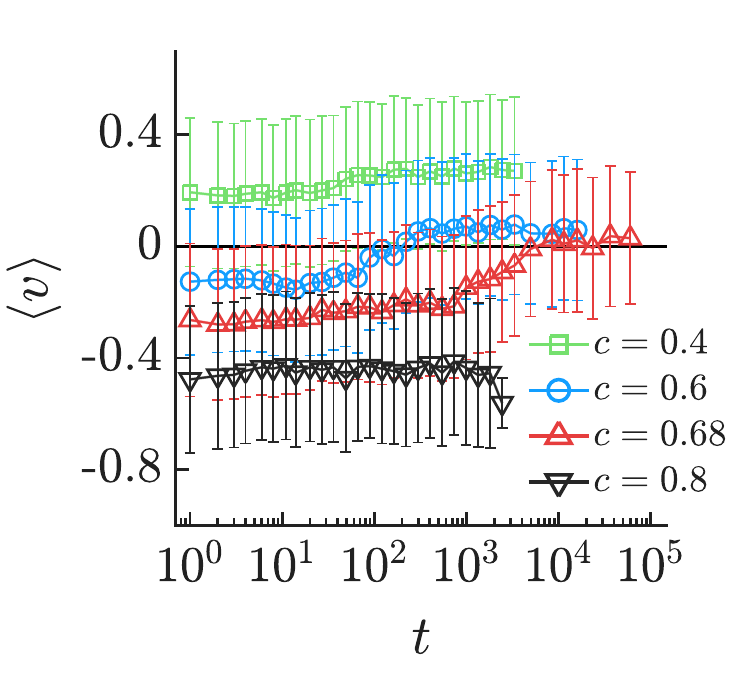}  
\caption{Time-evolution of the ensemble-averaged velocity of the C-D boundary for
  $c=0.4$, $c=0.6$, $0.68$, and $0.8$.}
\label{fig:boundary_velocity}
\end{figure}

The velocity of the C-D boundary,
defined as $v=x(t+1)-x(t)$ with $x(t)$ denoting the location of the rightmost cooperator and shown in Fig.~\ref{fig:boundary_velocity}, 
remains positive for $c=0.4$,
indicating persistent rightward motion until fixation to cooperation.
For $c=0.6$ and $0.68$, the velocity is initially negative but later becomes positive,
reflecting a reversal of boundary motion.
No such reversal occurs for $c=0.8$, where the velocity remains negative.  

\section{Influence of payoff-wealth ratio $\varepsilon$}
\label{sec:varepsilon}

\begin{figure}
\ff{.7}{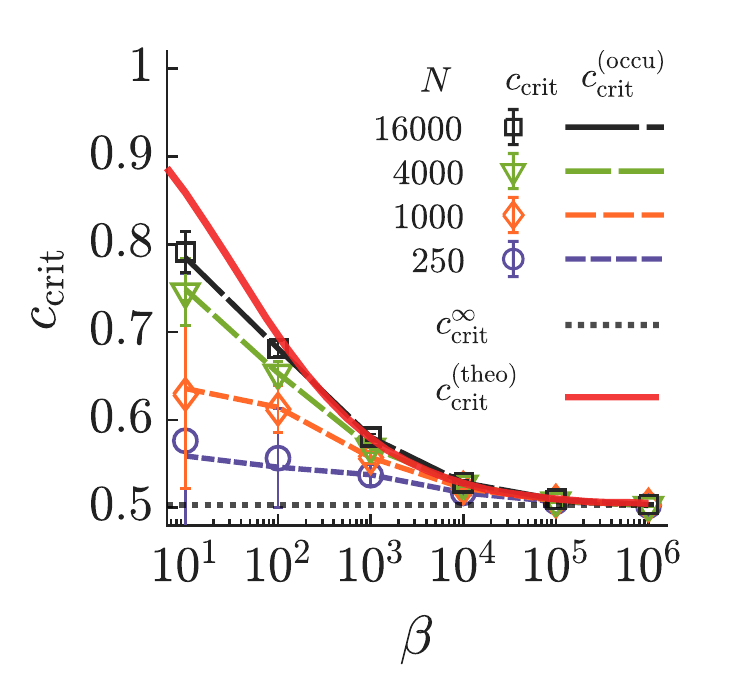}  
\caption{
  Critical points $c_\crit$ versus inverse temperature $\beta$ for
  $\varepsilon=0.001$ and different system sizes $N$.
  Shown are $c_\crit$ (points) and $c_\crit^\occupation$ (dashed) from simulations,
  $c_\crit^\theoretical$ (solid) from Eq.~\eqref{eq:self},
  and $c_\crit^{\infty}$  in the $\beta\to\infty$
  limit from Eq.~\eqref{eqn:beta_inf}.
  The error bars for $c_\crit$ were estimated as the range of $c$ values for
  which $0.2<\rho_\fixation<0.8$.
}
\label{fig:varepsilon}
\end{figure}

The constant $\varepsilon$ characterizes the ratio of payoff to wealth as formulated in
Eq. (1). It should be a small positive constant to prevent bankruptcy.
From Eq. (1) and the wealth update  $W_i(t+1) = W_i(t) + \sum_{j\in \mathcal{N}_i}\left(\sum_{k\in \mathcal{N}_i} +\sum_{k\in \mathcal{N}_j}\right) \Pi_{ij}^k(t)$ as in the main text,  $\varepsilon$ should satisfy $\varepsilon<{1\over 8d^2}$ in a $d$-dimensional lattice to ensure nonnegative wealth. This follows because the maximum possible decrease of $W_i$ between time $t$ and $t+1$ is $8 \varepsilon d^2 W_i(t)$, which must be smaller than $W_i(t)$ to prevent bankruptcy.  

Our main results are qualitatively insensitive to $\varepsilon$. We show in Fig.~\ref{fig:varepsilon} the critical point as a function of $\beta$ for different system sizes and $\varepsilon=0.001$, which is qualitatively similar to the case  $\varepsilon=0.01$ shown in Fig. 2(d). 

\section{Influence of temperature in two dimensions}
\label{sec:2d}

\begin{figure}[h]
\ff{0.8}{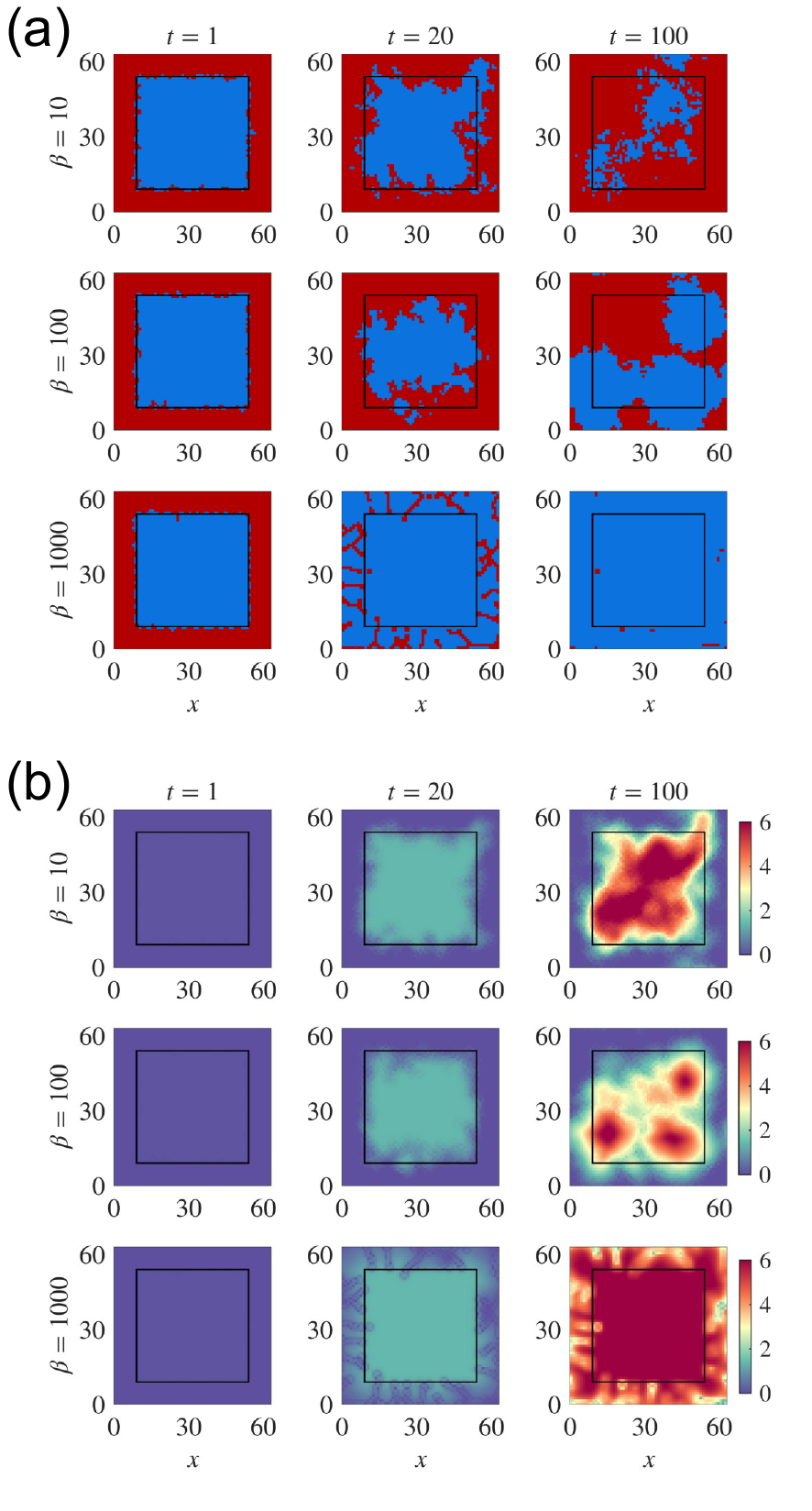}  
\caption{
  Time evolution of the strategy configuration and wealth profile on
  a two-dimensional lattice of size $N=64\times64$ with $c=0.45$ and $\varepsilon=0.01$.
  Snapshots are shown at inverse temperatures $\beta=10$, $100$, and $1000$,
  and at times $t=1$, $20$, and $100$.
  (a) Strategy configuration, blue denotes cooperators and red denotes defectors.
  (b) Wealth profile, colors indicate wealth on a $\log_{10}$ scale.
}
\label{fig:strategy_2d}
\end{figure}

The model rules of strategy evolution in the one-dimensional lattice naturally
extend to the two-dimensional lattice  of size $N=$ $ L \times L$.
We consider each node's four von Neumann neighborhoods,
and the periodic boundary conditions are imposed in both the $x$ and $y$ directions.
To supplement the two-dimensional results presented in the main text,
here we present snapshots of the simulations at different time steps for different values of
$\beta$ in Fig.~\ref{fig:strategy_2d}, which reveals significant $\beta$-dependence. 

At temperature increases ($\beta$ decreases),
spatial fragmentation of C-clusters becomes more frequent and severe,
eliminating small clusters and thereby lowering $c_\crit$.
Without wealth-scaled interactions, the critical point $c_\crit^\nowealth$ decreases with
increasing temperature,
demonstrating such  role of temperature in suppressing cooperation.
Even under the wealth-scaled interaction,
the C-clusters may perish under this negative influence of temperature
before sufficient boundary wealth  gradients are established to drive the  expansion of C-clusters.
Thus, the cooperation-suppressing effect of temperature competes with
its cooperation-enhancing effect under wealth-scaled interactions.
As shown in Fig. 3(b), this trade-off yields an optimal temperature for the cooperation
around $\beta \approx 3000$.

 \bibliography{References}
\end{document}